\documentclass[11pt]{article}

\usepackage[]{graphicx}
\usepackage{amsmath}
\usepackage{amssymb}
\usepackage{epsfig}

\usepackage{graphicx}
\begin{document}


\begin{titlepage}

\renewcommand{\thefootnote}{\alph{footnote}}
\vspace*{-3.cm}
\begin{flushright}

\end{flushright}

\vspace*{0.3in}

\renewcommand{\thefootnote}{\fnsymbol{footnote}}
\setcounter{footnote}{-1}

{\begin{center} {\Large\bf  The  effective resistance of the $N$-cycle graph with four nearest neighbors}
\end{center}}
\renewcommand{\thefootnote}{\alph{footnote}}

\vspace*{.8cm} {\begin{center} {\large{\sc
                }}
\end{center}}
\vspace*{0cm} {\it
\begin{center}
 \vspace*{.8cm} {\begin{center} {\large{\sc
                Noureddine~Chair
                }}
\end{center}}
\vspace*{0cm} {\it
\begin{center}
 Physics Department,
 The University of Jordan, Amman, Jordan

Email: n.chair@ju.edu.jo
\end{center}}
\end{center}}

\vspace*{1.5cm}

\begin{center} Abstract\end{center}\ \  The  exact expression for the  effective resistance between any two vertices of the $N$-cycle graph  with four nearest neighbors  $C_{N}(1,2)$,  is given. It turns out that this expression is written in terms of  the effective resistance of the $N$-cycle graph $C_{N}$,  the square of the Fibonacci numbers, and the bisected Fibonacci numbers.  As a consequence  closed form formulas for the total effective resistance, the first passage time (FPT), and the mean first passage time (MFPT) for the simple random walk on the the $N$-cycle graph with four nearest neighbors are obtained. Finally, a closed form formula for the effective resistance of  $C_{N}(1,2)$ with all first neighbors removed is obtained.     
 \vspace*{.5cm}
\end{titlepage}

\renewcommand{\thefootnote}{\arabic{footnote}}
\setcounter{footnote}{0}


\section{Introduction}\ 
A classical problem in electric circuit is the computation  of the resistance between two nodes in a resistor network (some relevant references  may be found in \cite{Cserti})
. Calculating the resistance using the traditional analysis methods,  such as  the Kirchhoff's laws, can be done in principle  for any given network, however, with the growing
number of nodes, the number of equations to solve growths very rapidly. Most of resistance computations in the past are carried out for infinite networks using the Green’s function technique \cite{Cserti, Cserti1}.

\ \ In 2004, Wu \cite{ Wu} proposed a new formulation of  finite  resistor networks that gives a formula  for  the two-point resistance (effective resistance) in terms of the eigenvalues and the eigenvectors of the Laplacian matrix. This formula may be used to compute the effective resistance of the  one-dimensional lattice, two-dimensional network and higher-dimensional lattices. Known results for infinite networks may be obtained by taking the infinite-size limit \cite{Cserti}.\\
It has been known for sometime now,  that  electrical networks and random walks are closely related. They are connected through the effective resistance of an electrical network.  One of the interesting  parameter that characterizes  and give information about  random walks is  the first passage time ( FPT ), or the hitting time,  this is the expected time to hit a target node for the first time  for a walker starting from a source node.  Also, other quantities of interests to know  are  the mean first passage time (MFPT) and  the commute time (the random round tripe between two nodes) \cite{Doyle, Tetali, Chandra}. The close link between the effective resistance and the commute time of a random walker
on the graph was established in \cite{ Chandra}. 
 Therefore, the effective resistance provides an alternative way to compute the FPT.
 
 \ \  In this paper, the resistor electrical network  is a connected, undirected graph $G=(V, E)$, with  vertex set $ V=(1,2,.\cdots N)$ and  edge set  $E$ in which each edge $ij$ has a
resistance $r_{xy}=1$.  An interesting invariant quantity related to the effective  resistance  of a graph $G$, is the  total effective resistance of $G$, denoted by  $ R(G)$ \cite{randic},  it is  called the Kirchhoff's index  this is the analogue of the Wiener index $W(G)$  of a graph $G$ \cite{ Wiener}. This quantity tuns out to be connected to the MFPT.

\ \   For a given graph $G$, it is not always possible to get closed formula for the effective  resistance of graphs, however, for  certain graphs with symmetries like the undirected circulant graphs \cite{Biggs}, this may be possible.  By using spectral graph theory \cite{Lovasz}, Chau and Basu  \cite{ Chau}  obtained a formula to compute the FPT of the random walk on the $N$-cycle graph with $2p$ neighbors. However, We find that in this case, it is easier  to use  Wu's formula to obtain the effective resistance between any two nodes. This follows from 
the fact that the eigenvalues of the Laplacian matrix may be obtained without any difficulty in this case. Then it is not difficult to show that this formula  when multiplied by the number of  edges $\vert E\vert  $  is identical to the (FPT) given in \cite{ Chau}. The undirected circulant graphs  enjoy rotational symmetry, i.e., each vertex of these  graphs  ‘looks the same’ from any vertex. Therefore, we may as well consider that the random walk has  started at vertex $0$,  and after some steps reaches a given vertex say $l$.

\ \ Using the commute time  formula given by  Chandra et al.  \cite{Chandra}, $ C_{ij }=2 \vert E\vert R_{ij}$, then  the first passage time  $H_{0l }$ may be written as  $H_{0,l }=\vert E\vert R_{0,l} $. For example, the effective resistance between the vertex $0$ and any other vertex $l$ of the complete graph is $ R_{0,l}= 2/N$, and since  the number of edges is $ \vert E\vert=\frac{N(N-1)}{2}$, then the expression for the FPT of the random walk on the complete graph gives   $H_{0,l }=N-1$. This result  was  derived previously using probabilistic techniques  on graphs \cite{ Lovasz, Aldous}.

\ \ It was pointed out by Wu \cite{Wu},  that little attention has been paid to finite electrical network, even though these are the ones occurring in real life. Therefore, it interesting to apply such a formalism to other finite electrical networks beyond the cycle and the complete graphs \cite{Wu}.  We have already done so  for the complete graph $K_{N}$ in which the number of vertices $N$ is odd, minus the $N$ edges that connect the opposite vertices  \cite{Chair1}. Here, we give an  exact expression for  the  effective resistance between any two vertices of the $N$-cycle graph  with four nearest neighbors, i.e., every vertex is connected to its two neighbors and neighbor's neighbors, this graph is denoted by $C_{N}(1,2)$.

\ \   The formalism  developed in this paper and in \cite{Chair1}, enables us to obtain closed formulas for the effective resistance of some finite two-dimensional electrical networks, like The circular ladder and the $2\times N$ electrical network with free boundaries \cite{Chair}.
 It turns out  that the effective resistance of the  graph $C_{N}(1,2)$ with its first neighbors being removed  contributes  to the effective resistance of the circular ladder.
 
 \ \ In general, the effective resistance of the circulant graphs are given by trigonometrical sums, in particular, when considering  the graph  $C_{N}(1,2)$, one has to deal with trigonometrical power  sums. This  may be evaluated  using  a  formula by Schwatt \cite{ Schwatt} on trigonometrical power sums, however, this formula does not give the right answer when the powers are congruent to $N$. Therefore, one has to modify  the formula  slightly to get the right answer.
 
 \ \ Here,  the exact computations  of the effective resistance and the total effective resistance are  explicit, this is done in Section 2. In Section 3, the the FPT, and the MFPT   of the simple random walks on the graph $C_{N}(1,2)$ are also given by closed formulas. Computation  of the effective resistance of the graph $C_{N}(1,2)$ minus $2N$ vertices of the first neighbors  is considered in Section 4. Finally, 
in section 5, we give our discussion and check our formulas using the Foster first identity \cite{Foster} and the recursion formula satisfied by the hitting time. 
\section {The Exact effective resistance and the total  effective resistance of the graph $C_{N}(1,2)$ }\ \
Here, we give the exact formula for the  effective resistance of the  $N$-cycle graph with four nearest neighbors, see Figure. 1.  According to  Wu's  formula \cite{Wu}, given a  resistor network with unit resistors,  the effective resistance between any two nodes $\alpha$ and $\beta$  is given by 
\begin{equation}
\label{7}
R_{\alpha,\beta}=\sum_{n=1}^{N-1}\frac{|\psi_{n\alpha}-\psi_{n\beta}|^2}{\lambda_n},
\end{equation} 
where $1\leq\alpha,\beta \leq N$ and $\lambda_n$,  $\psi_{n}$ are the eigenvalues and the eigenvectors of the  Laplacian $L$ of the resistor network having unit resistors (edges), that is, graphs.

\ \  The Laplacian  matrix $L$ of  the graph  $G=(V,E)$ is  $ L=D-A$, here, $D$ is the diagonal matrix of degrees whose   $i^{\text{ith}}$ element  is  the number of resistors connected to the $i^{\text{ith}}$ node. And $ A$ is the adjacency matrix  representing the edge set $E$.

\ \ The matrix elements of the Laplacian of the graph  $C_{N}(1,2)$ are;
\begin{eqnarray}
L_{mn}= \left\{\begin{array}{rl}
4 & \text{if } m=n,\\
-1 & \text{if } m=n\pm1, n\pm2 \mod(N),\\
0  & \text{otherwise}, 
\end{array} \right.
\end{eqnarray}
equivalently, we may writes  $ L_{mn}=4\delta_{m,n}-(\delta_{m,n+1}+\delta_{m,n-1})-(\delta_{m,n+2}+\delta_{m,n-2})$.

\ \ The laplacian matrix $L$ may be diagonalized using the basis $ \psi_{n,k}:=\frac{1}{\sqrt{N}}\exp(\frac{2\pi ink}{N})$. To that end, we make use of the fact that the matrix elements $\delta_{m,n+ k} $ may be considered as matrix elements of the   $k^{\text{th}}$ power of the rotation matrix $R$ of finite  closed lattice \cite{Wolf}, i.e, $(R^{k})_{mn}=\delta_{m,n+ k}$. Using the expression of the matrix  $\psi$, then  $\delta_{m,n+ k} $ may be written as
\begin{eqnarray}
\label{a5}
(\mathcal R^{k})_{mn}&=&(\psi^{*}R^{k}\psi)_{mn}=\frac{1}{N}\sum_{j,l} \delta_{j,l+ k} \exp[2\pi i(ln-jm)/N] \nonumber\\&=&\exp[-2\pi ikm/N]\frac{1}{N}\sum_{l}\exp[2\pi il(n-m)/N]\nonumber\\&=& \delta_{m,n}\exp[-2\pi ikm/N],
\end{eqnarray}
similarly one shows $(R^{-k})_{mn}=\delta_{m,n- k}$ in the diagonalized form reads; $$ (\mathcal R^{-k})_{mn}=\delta_{m,n}\exp[2\pi ikm/N].$$.

\ \  As a consequence, the eigenvalues of the laplacian of the $N$-cycle graph with four nearest neighbors,  are   $ \lambda_n=4\sin^2(n\pi/N)+4\sin^2(2n\pi/N)$ . Thus,  the effective  resistance of the  graph $C_{N}(1,2)$ is 
\begin{eqnarray} 
\label{sum1}
R(l) &=&\frac{1}{N}\sum_{n=1}^{N-1}\frac{4\sin^2(nl\pi/N)}{4\sin^2(n\pi/N)+4\sin^2(2n\pi/N)}\nonumber \\ &=&\frac{1}{5N}\sum_{n=1}^{N-1}\frac{\sin^2(nl\pi/N)}{\sin^2(n\pi/N)}+\frac{4}{5N}\sum_{n=1}^{N-1}\frac{\sin^2(nl\pi/N)}{5(1-4/5\sin^2(n\pi/N)},
\end{eqnarray}
up to a constant factor, the first term in the above  expression  is nothing but the effective resistance of the $N$-cycle graph \cite{Wu},
 \begin{eqnarray} 
\label{Wu}
\frac{1}{N}\sum_{n=1}^{N-1}\frac{\sin^2(nl\pi/N)}{\sin^2(n\pi/N)}=l\big(1-l/N\big),
\end{eqnarray} 
 while the second trigonometrical sum is a non-trivial sum that we want to evaluate.
 
 \ \  This term may be written explicitly as;
\begin{eqnarray}
\label{Cont}
\sum_{n=1}^{N-1}\frac{\sin^2(nl\pi/N)}{(1-4/5\sin^2(n\pi/N)}=\frac{1}{2}\sum_{j=0}^{\infty}(4/5)^{j}\Big\lbrack\sum_{s=1}^{l}(-1)^{s+1} \frac{l}{l+s}  \binom {l+s} {l-s}2^{2s}\sum_{n=1}^{N-1}\sin^{2(s+j)}(n\pi/N)\Big\rbrack,
\end{eqnarray} 
in obtainig Eq. (\ref{Cont}),  we  used the identity $ \cos2(ln\pi/N)=\sum_{s=0}^{l}(-1)^s \frac{l}{l+s}\binom {l+s} {l-s}2^{2s}\sin^{2s}(n\pi/N)$.  The  sum over $n$ in  Eq. (\ref{Cont}), may be found in  Schwatt's classic book \cite{Schwatt}, however, it is not suitable for  $j$  a multiple of $N$, so his formula must  be modified . We may show that the right formula that takes into consideration this fact is
\begin{eqnarray}
\label{trig1}
\sum_{n=1}^{N-1}\sin^{2(s+j)}(n\pi/N)=
\frac{N}{2^{2(s+j)}}\binom {2(j+s)}{j+s} +\frac{N}{2^{2(j+s)-1}}\sum_{p=1}^{[(j+s)/N]}(-1)^{pN}\binom {2(j+s)} {j-pN}.
\end{eqnarray}\ \
 Fortunately, we do not need to have a closed form formula for the binomial sum to  do  our  computations. This problem is by-passed through the residue representation of binomials, this will be discussed shortly.
 Let  $R_{1}(l)$ be the above second term of Eq. (\ref{sum1}), then our computation  is equivalent to evaluate the following  terms  
\begin{eqnarray}
\label{trig2}
R_{1}(l)^{'}=\frac{1}{25}\sum_{s=1}^{l}(-1)^{s+1} 5^{s} \frac{2l}{l+s}\binom {l+s} {l-s}\Big\lbrack\sum_{J=0}^{\infty}(1/5)^{J}\binom {2J} {J}-\sum_{J=0}^{s-1}(1/5)^{J}\binom {2J} {J}\Big\rbrack,
\end{eqnarray}
and
\begin{eqnarray}
\label{trig3}
R_{1}(l)^{''}=\frac{4}{25}\sum_{J=0}^{\infty}(1/5)^{J-5}\sum_{s=1}^{l}(-1)^{s+1}  \frac{l}{l+s} \binom {l+s} {l-s}\sum_{p=1}^{[J/N]}(-1)^{pN}\binom  {2J} {J-pN},
\end{eqnarray}
where $J=j+s$, note that in obtaining the above second  sum over $J$, we added the terms $J=0,\cdots J=s-1$, since  they do not contribute to the binomial $\binom {2J} {J-pN} $, and $ l\leq{N-1}$.

\ \ Now, we will  evaluate  the sum over $J$, this is done through  the representation of  binomial coefficients by residue \cite{Egorychev}.
First, recall if $ G(w)$ is the  generating function of the binomial coefficient sequence  $\binom {n}{k}$ for a fixed $n$  given by $G(w)= \sum_{k=0}^{n}\binom {n}{k} w^{k} =(1+w)^{n},$ 
then  $\binom {n}{k}=\hbox{res}_w (1+w){^n}{w^{-k-1}}$. The following linearity property of the residue operator $\hbox{res}$ is very  crucial in evaluating the effective resistance. This may be defined as follows;  given some constants $\alpha$ and $\beta$, then $\alpha \hbox{res}_w G_{1}(w){w^{-k-1}}+ \beta\hbox{res}_w G_{2}(w){w^{-k-1}}=\hbox{res}_w(\alpha G_{1}(w)+ \beta G_{2}(w)) {w^{-k-1}}.$ Using the residue representation of the binomials and the linearity  property, then the sum over $J$ in Eq. (\ref{trig2}) gives
 \begin{equation}
\label{trig4}
R_{1}(l)^{'}=\frac{1}{25}\sum_{s=1}^{l}(-1)^{s+1} 5^{s} \frac{2l}{l+s}\binom {l+s} {l-s}\Bigg\lbrack \sqrt{5}-5^{1-s}\hbox{res}_w\Bigg(\frac{(1+w)^{2s}}{(w){^s}((1+w)^2-5w)}\bigg)\Bigg\rbrack.
\end{equation}\ \
The first sum in Eq. (\ref{trig4}) may be written in terms of the Lucas numbers through the normalized  Chebyshev polynomial of the first kind  \cite{Rivlin}. Let this  sum be denoted by $L_{1}(l)$, then
\begin{equation}
\label{trig5}
L_{1}(l)=(-1)^{l+1}\frac{\sqrt{5}}{25}\sum_{k=0}^{l-1}(-1)^{k}5^{l-k}  \frac{2l}{2l-k}\binom {2l-k} {k}=(-1)^{l+1}\frac{\sqrt{5}}{25}\Bigg\lbrack C_{2l}(\sqrt{5}) -(-1)^{l}2\Bigg\rbrack ,
\end{equation}
where $$ C_{2l}(x)= 2T_{2l}(x/2)=\sum_{k=0}^{l}(-1)^{k} \frac{2l}{2l-k}\binom {2l-k} {k}x^{2l-2k}$$ is the normalized Chebyshev polynomial of the first kind  and
$$T_{2l}(x/2)=\frac{1}{2}\Bigg\lbrack\Bigg(\frac{x}{2}+\sqrt{(x/2)^2-1}\bigg)^{2l} +\Bigg(\frac{x}{2}-\sqrt{(x/2)^2-1}\bigg)^{2l} \Bigg\rbrack.$$ 
Thus, $$L_{1}(l)=(-1)^{l+1}\frac{\sqrt{5}}{25}\bigg(L_{2l}-2(-1)^{l} \bigg)
 =(-1)^{l+1}\frac{\sqrt{5}} {5} F_{l}^{2},
$$
where $$ L_{2l}=\Bigg(\frac{1+\sqrt{5}}{2}\bigg)^{2l}+\Bigg(\frac{1-\sqrt{5}}{2}\bigg)^{2l},$$ are the bisection of the Lucas numbers, and  $$F_{l}=\frac{1}{\sqrt{5}}\Big(\frac{1+\sqrt{5}}{2}\bigg)^{l}-\frac{1}{\sqrt{5}}\Big(\frac{1-\sqrt{5}}{2}\bigg)^{l},$$ are the well-known Fibonacci numbers.

\ \ The second sum in  Eq. (\ref{trig4}) denoted by $F_{1}(l)$,  may be computed to give
 \begin{eqnarray}
 \label{trig6}
 F_{1}(l)&=&(-1)^{l}\frac{1}{5}\hbox{res}_{w=0}\frac{1}{(w-(3-\sqrt{5})/2)(w-(3+\sqrt{5})/2))}\bigg({C_{2l}({\frac{1}{\sqrt{w}}+\sqrt{w}})-(-1)^{l}2}\bigg)\nonumber\\
&=&(-1)^{l+1}\frac{1}{5\sqrt{5}}\hbox{res}_{w=0}\frac{1}{w^{l}}\bigg(\frac{1}{((w-(3-\sqrt{5})/2))}-\frac{1}{((w-(3+\sqrt{5})/2))}\bigg)\nonumber\\
&=&(-1)^{l}\frac{1}{5}F_{2l}.
 \end{eqnarray}
 Thus, the sum  $R_{1}(l)^{'}$ is given by  the following  closed form formula 
\begin{equation}
\label{trig7}
R_{1}(l)^{'}=(-1)^{l+1}\frac{\sqrt{5}} {5} F_{l}^{2}+(-1)^{l}\frac{1}{5}F_{2l}
\end{equation}\ \
For  $N$  even,  the sum $R_{1}(l)^{''}$ given in Eq. ( \ref{trig3} ) is
\begin{eqnarray}
\label{trig8}
R_{1}(l)^{''}=\frac{4}{25}\sum_{J=0}^{\infty}(1/5)^{J-5}\sum_{s=1}^{l}(-1)^{s+1}  \frac{l}{l+s} \binom {l+s} {l-s}\sum_{p=1}^{[J/N]}\binom  {2J} {J-pN},
\end{eqnarray}
where 
 \begin{equation}
 \label{trig12}
 \sum_{p=1}^{[J/N]}\binom  {2J} {J-pN}=\hbox{res}_{w}(1+w)^{2J}w^{-J-1}\Big(\frac{w^{N}}{1-w^{N}} \Big).
 \end{equation}
Then the  second term  $R_{1}(l)^{''}$, becomes
\begin{eqnarray}
\label{trig13}
R_{1}(l)^{''}=\frac{2}{5}\sum_{s=1}^{l}(-1)^{s+1}5^{s}  \frac{2l}{l+s} \binom {l+s} {l-s} \hbox{res}_{w}\Big(\frac{w^{N}}{1-w^{N}}\frac{1}{5w-(1+w)^{2} }\Big),
\end{eqnarray}
the acceptable residue is $ w=(3-\sqrt{5})/2$, and so the expression for $R_{1}(l)^{''}$ is
\begin{eqnarray}
\label{trig14}
R_{1}(l)^{''}=\frac{2}{5\sqrt{5}}\sum_{s=1}^{l}(-1)^{s+1}5^{s} \frac{2l}{l+s} \binom {l+s} {l-s}\Bigg(\frac{(\frac{3-\sqrt{5}}{2})^{N}}{1-(\frac{3-\sqrt{5}}{2})^{N}}\Bigg).
\end{eqnarray}\ \
The sum over $s$, was  done previously, see Eq. (\ref{trig5}), therefore, 
\begin{eqnarray}
\label{trig15}
R_{1}(l)^{''}=(-1)^{l+1}\frac{2} {\sqrt{5}} F_{l}^{2}\Bigg(\frac{(\frac{3-\sqrt{5}}{2})^{N}}{1-(\frac{3-\sqrt{5}}{2})^{N}}\Bigg).
\end{eqnarray}\ \
Finally, the exact formula for the effective  resistance between any two vertices of the  graph $ C_{N}(1,2)$, reads
\begin{eqnarray}
\label{trig16}
R(l)&=&\frac{1}{5}{l}(1-l/N)+(-1)^{l+1}\frac{\sqrt{5}}{5}F_{l}^{2}+(-1)^{l}\frac{1}{5}F_{2l}\nonumber\\&+&
(-1)^{l+1}\frac{2}{\sqrt{5}}F_{l}^{2}\Bigg(\frac{(\frac{3-\sqrt{5}}{2})^{N}}{1-(\frac{3-\sqrt{5}}{2})^{N}}\Bigg)\nonumber\\&=&\frac{1}{5}l(1-l/N)+(-1)^{l+1}\frac{ F_{l}^{2}}{\sqrt{5}}\Bigg(\frac{1+(\frac{3-\sqrt{5}}{2})^{N}}{1-(\frac{3-\sqrt{5}}{2})^{N}}\Bigg)+(-1)^{l}\frac{ F_{2l}}{5}.
\end{eqnarray}\ \
Our formula is consistent with the recursion relation, 
$R(l+1)=l(1-l/N)+2R(1)-3R(l)-R(l-1)$.
The effective resistance  $R(l)$ given by Eq. (\ref{sum1}) enjoys the following symmetry  $R(l)=R(N-l) $, this is obvious from the first term of our formula given in Eq.  (\ref{trig16}), however, from the second term, this symmetry is not obvious. If our formula were to be correct, then the symmetry enjoyed by the effective resistance $R(l)$ would give the following identity,
\begin{eqnarray}
\label{trig18}
\frac{1}{5}\Big( F_{2(N-l)}- F_{2l}\Big)=\frac{1}{\sqrt{5}}\Big(F_{(N-l)}^{2}-F_{l}^{2}\Big)\Bigg(\frac{1+(\frac{3-\sqrt{5}}{2})^{N}}{1-(\frac{3-\sqrt{5}}{2})^{N}}\Bigg).
\end{eqnarray}
This turns out to be  correct and follows simply by using the different expressions of the Fibonacci numbers.

\ \ The effective resistance considered so far, assumes that the number of vertices is even, the odd case, however, is not so different as far as the above computations are  concerned, the only difference between the two cases is that  the sum over $p$ of the binomial given by Eq. (\ref{trig12}), should be replaced by
\begin{equation}
 \label{trig19}
 \sum_{p=1}^{[J/N]}(-1)^{p}\binom  {2J} {J-pN}=-\hbox{res}_{w}(1+w)^{2J}w^{-J-1}\Big(\frac{w^{N}}{1+w^{N}} \Big).
 \end{equation}
 Therefore, the exact effective resistance of the graph $ C_{N}(1,2)$, for $N$  odd reads
 \begin{equation}
\label{trig20}
R(l)=\frac{1}{5}\frac{l}{N}(N-l)+(-1)^{l+1}\frac{ F_{l}^{2}}{\sqrt{5}}\Bigg(\frac{1-(\frac{3-\sqrt{5}}{2})^{N}}{1+(\frac{3-\sqrt{5}}{2})^{N}}\Bigg)+(-1)^{l}\frac{ F_{2l}}{5}.
\end {equation}
The symmetry of the effective resistance in this case implies the following identity 
\begin{eqnarray}
\label{trig21}
\frac{1}{5}\Big( F_{2(N-l)}+ F_{2l}\Big)=\frac{1}{\sqrt{5}}\Big(F_{(N-l)}^{2}+F_{l}^{2}\Big)\Bigg(\frac{1-(\frac{3-\sqrt{5}}{2})^{N}}{1+(\frac{3-\sqrt{5}}{2})^{N}}\Bigg).
\end{eqnarray}\ \
Therefore, the computed effective resistance $R(l)$ is correct and satisfies the hidden symmetry property $R(l)=R(N-l) $.

\ \ Having computed the effective resistance of the  graph $C_{N}(1,2)$, we now compute the total effective resistance of  $C_{N}(1,2)$, this is also known as the  Kirchhoff index \cite{randic}. The total effective resistance  of  $C_{N}(1,2)$ in terms of the effective resistance  $R(l)$,  is
$$R(C_{N}(1,2))=\sum_{i<j}R_{ij}=\frac{N}{2}\sum_{l=1}^{N-1}R(l).$$

\ \  In order to sum over all the effective resistances  given by the formula  (\ref{trig16}) for $N$ even, one needs to evaluate the alternating  sums  $\sum_{l=1}^{N-1}(-1)^{l} F_{l}^{2} $ and  $\sum_{l=1}^{N-1}(-1)^{l} F_{2l} $. The first alternating sum may be obtained using the identity due to Lucas $ L_{2l}= F_{2l+2}- F_{2l-2} $, to give
$$\sum_{l=1}^{N-1}(-1)^{l} F_{l}^{2}=-\frac{1}{5}\big( F_{2N-1}-1+2N\big).$$ 
Similarly, the second alternating sum may be evaluated from another Lucas formula $  F_{2l}=F_{l+1}^{2}-F_{l-1}^{2}$, to give
$$\sum_{l=1}^{N-1}(-1)^{l} F_{2l}=  F_{N-1}^{2}-F_{N}^{2}-1 .$$\ \
Therefore the exact formula for the total effective resistance reads
\begin{eqnarray}
\label{trig22}
R(C_{N}(1,2))&=&\frac{1}{5}\frac{N^3-N}{12}+\frac{N}{2}\frac{1}{5}\big( F_{2N-1}+2N-1\big)\frac{ 1}{\sqrt{5}}\Bigg(\frac{1+(\frac{3-\sqrt{5}}{2})^{N}}{1-(\frac{3-\sqrt{5}}{2})^{N}}\Bigg)\nonumber\\&+&\frac{N}{2}\frac{1}{5}\big(F_{N-1}^{2}-F_{N}^{2}-1\big).
\end{eqnarray}\ \ 
For $N$ odd,  the total effective resistance is 
\begin{eqnarray}
\label{trig23}
R(C_{N}(1,2))&=&\frac{1}{5}\frac{N^3-N}{12}-\frac{N}{2}\frac{1}{5}\big( F_{2N-1}-2N+1\big)\frac{ 1}{\sqrt{5}}\Bigg(\frac{1-(\frac{3-\sqrt{5}}{2})^{N}}{1+(\frac{3-\sqrt{5}}{2})^{N}}\Bigg)\nonumber\\&+&\frac{N}{2}\frac{1}{5}\big(F_{N}^{2}-F_{N-1}^{2}-1\big).
\end{eqnarray}
\section{ Simple random walk on the graph $C_{N}(1,2) $}\ \
The connection between simple random walks and  electrical networks  via the effective resistances in  electrical networks is well  established \cite{Doyle, Tetali, Chandra}. In particular, the covering and commute times of random walks in graphs  can be determined by the
effective resistance  \cite{Tetali, Chandra}. The simple random walk  on a graph $G$ is defined by the jumping probability $p_{ij}$ between nearest neighbor vertices $i$ and $j$:
\begin{eqnarray}
 p_{ij}=\left\{\begin{array}{cl} \frac{1}{d(i)}& \text{ if  $i$, $j$ {are adjacent vertices}}, \\
0& \ \text { otherwise}\
\end{array}\right.
\end{eqnarray}
here,  $d(i)$ is the degree of the vertex $i$.

\ \  
This may be interpreted as  the probability the walk  move from vertex $i$ to  vertex $j$, given that we are at vertex $i$. The simple random walk on a graph $G$, may be described  by the first passage time, or the hitting time FPT, $ H_{ij}$, this is the expected number of jumps that the walk takes until it lands on $j$, assuming the walk starts at $i$.

\ \ The other related parameter is the the commute time $C_{ij}$, this is the expected number of jumps in a random walk starting at $i$ before vertex  $j$ and then vertex $i$ is reached again, that is,  $ C_{ij}=H_{ij}+H_{ji}$. The discussion of hitting times in terms of the effective resistance seems to be  due to Chandra et al  \cite{ Chandra}. In particular,  it has been proved by Chandra et al,  that the commute time $ C_{ij }$ is equal to $2 \vert E\vert R_{ij}$.

\ \  It has been shown in \cite{Gutman}, that the total effective resistance of a graph $G$,  $R(G)=\sum_{i<j}R_{ij}$, may be written as
$$   R(G)=N\sum_{n=1}^{N-1}\frac{1}{\lambda_n},$$  where $\lambda_{n}$ are the eigenvalues of the Laplacian of a graph $G$. Thus, the total effective resistance is an invariant quantity of a graph $G$, for symmetric graphs like the circulant graphs with  the property $H_{ij}=H_{ji }$,  this quantity reads,
 \begin{equation}
\label{com1} R(G)=\frac{1}{2\vert E\vert}\sum_{i<j}(H_{ij}+H_{ji})=\frac{1}{\vert E\vert}\sum_{i<j}H_{ij}
\end{equation}
that is, the total effective resistance of the symmetric graph $G$, is the hitting time  averaged over all pairs of vertices.

\ \  If the expression for the  effective resistance is known in a closed form, then so are the total effective resistance, the FPT and MFPT. Since the  graph $C_{N}(1,2)$ has rotational symmetry, that is, the FPT may be written as  $H_{0,l}=\vert E\vert R(l)$, and with the total number of edges  $\vert E\vert=2N$,  then, the  exact expression for the FPT of the simple random walk on the $N$-cycle graph with $4$-nearest neighbors is 
 \begin{eqnarray}
\label{first hit}
H_{0,l}= \left\{\begin{array}{cl}
\frac{2}{5}l(N-l)+(-1)^{l+1}\frac{2N}{\sqrt{5}} F_{l}^{2}\Bigg(\frac{1+(\frac{3-\sqrt{5}}{2})^{N}}{1-(\frac{3-\sqrt{5}}{2})^{N}}\Bigg)+(-1)^{l}\frac{ 2N}{5}F_{2l}  & \text{for $N$ even },\\
\frac{2}{5}l(N-l)+(-1)^{l+1}\frac{2N}{\sqrt{5}} F_{l}^{2}\Bigg(\frac{1-(\frac{3-\sqrt{5}}{2})^{N}}{1+(\frac{3-\sqrt{5}}{2})^{N}}\Bigg)+(-1)^{l}\frac{ 2N}{5}F_{2l} & \text{for $N $ odd }
\end{array} \right.
\end{eqnarray}\ \
One may check that our expressions for the effective resistance of the $N$-cycle with $2p$  neighbors, and hence that of the first passage time agree with those in \cite{Chau}. This may be checked as follows;  the eigenvalues of the laplacian of the $N$-cycle graph with $2p$ neighbors  are;  $ \lambda_n=4\sum_{m1}^{p}\sin^2(nm\pi/N)$. Thus,  the two point resistance formula may be written as $$ R(l)=\frac{1}{N}\sum_{n=1}^{N-1}\sin^2(nl\pi/N){\Bigl{(}\sum_{m=1}^{p}\sin^2mn\pi/N}\Bigr{)}^{-1}.$$ The first passage time $H_{0,l}$ is $\vert E\vert  R(l) $, where the number of edges for the $N$-cycle with $2p$ neighbors is $\vert E\vert =pN$, and so $$H_{0,l}= p\sum_{n=1}^{N-1}\sin^2(nl\pi/N){\Bigl{(}\sum_{m=1}^{p}\sin^2mn\pi/N}\Bigr{)}^{-1}.$$ This is exactly equivlent to the formula given in \cite{Chau}.

\ \  The MFPT, may be obtained  from the following simple formula that holds for all regular graphs, i.e.,  all vertices that have the same degree $d$. This formula may be derived as follows; 
 \begin{equation} 
 \label{mean first hit}
 \overline{H_{0,l}}=\frac{1}{N}\sum_{l=1}^{N-1} H_{0,l} =\frac{2\vert E\vert}{N^2}\sum_{l=1}^{N-1}\frac{N}{2}R(l)=\frac{d}{N}R(G),
\end{equation}
 where we have used the formula $2\vert E\vert= Nd$ for regular graphs. Since $d=4$ for $C_{N}(1,2)$, the closed formula for the MFPT is
\begin{eqnarray}
\label{mean first hit1}
\overline{H_{0,l}} = \left\{\begin{array} {cl}
\frac{4}{5}\frac{N^2-1}{12}+\frac{2}{5}\big( F_{2N-1}+2N-1\big)\frac{ 1}{\sqrt{5}}\Bigg(\frac{1+(\frac{3-\sqrt{5}}{2})^{N}}{1-(\frac{3-\sqrt{5}}{2})^{N}}\Bigg)
+\frac{2}{5}\big(F_{N-1}^{2}-F_{N}^{2}-1\big)& \text{for $N$ even }\\
\frac{4}{5}\frac{N^2-1}{12}-\frac{2}{5}\big( F_{2N-1}-2N+1\big)\frac{ 1}{\sqrt{5}}\Bigg(\frac{1-(\frac{3-\sqrt{5}}{2})^{N}}{1+(\frac{3-\sqrt{5}}{2})^{N}}\Bigg)+\frac{2}{5}\big(F_{N}^{2}-F_{N-1}^{2}-1\big)& \text{for $N $ odd }
\end{array} \right.
\end{eqnarray}\ \
By using the formula $R(G)=N\sum_{n=1}^{N-1}\frac{1}{\lambda_n}$,  then the equivalent formula for the MFPT is 
\begin{equation}
\label{mean hit equi}
\overline{H_{0,l}}=\sum_{n=1}^{N-1}\frac{1}{\sin^2(n\pi/N)+\sin^2(2n\pi/N)}.
\end{equation}
This is the eigentime identity relating the mean first passage time to the eigenvalues of the Laplacian of the regular graph $C_{N}(1,2)$ of degree $4$.

\ \  In the literature, the eigentime identity is given  in terms of the eigenvalues of the normalized Laplacian $ \mathcal{L}$  \cite{Lovasz}, for regular graphs one has, $ \mathcal{L}=D^{-1}L$, i.e, the eigentime identity is independent of the Laplacian used. It is not difficult to show that the formula to compute the  MFPT between vertex $ 0$ and vertex $l$, on the $N$-cycle graph with $2p$ neighbors, is $$\overline{H_{0,l}} =p\sum_{n=1}^{N-1}{\Bigl{(}2\sum_{m=1}^{p}\sin^2mn\pi/N}\Bigr{)}^{-1}, $$
for $p=1$, gives the well known eigentime identity on the $N$-cycle \cite{Aldous}, and for $p=2$ gives Eq.(\ref{mean hit equi}).
\section{The graph $C_{N}(1,2)$ minus $2N$ vertices}\ \
Having obtained a closed formula for the effective resistance of the graph $C_{N}(1,2)$, next  we  compute  the two-point resistance of the graph  $C_{N}(1,2)$ minus $2N$ vertices, the total number  of  adjacent vertices in  the $N$-cycle graph.  Let us denote  this graph by $C_{N}^{-2N}(1,2)$, the eigenvalues of this graph are given by $\lambda_{n}=\sin^22n\pi/N$, then the effective resistance may be written as 
\begin{eqnarray}
\label{aa}
R(l)=\frac{1}{N}\sum_{n=1}^{N-1}\frac{\sin^2nl\pi/N}{\sin^22n\pi/N}.
\end{eqnarray}\ \
Before computing the effective resistance,   we make  the following observations,  for $N$  even,  the graph   $C_{N}^{-2N}(1,2)$  splits into two $N/2$-cycles, see Figure. 2. Therefore,  the effective resistance for $N$ even reeds; 
\begin{eqnarray}
\label{aaaa}
R(l)=\frac{2}{N}\sum_{n=1}^{N/2-1}\frac{\sin^22nl\pi/N}{\sin^22n\pi/N}=l(1-2l/N),
\end{eqnarray}
in obtaining the above equation we used Wu's formula Eq. (\ref{Wu}).

\ \  
Now, if $N$ is odd, then the graph  $C_{N}^{-2N}(1,2)$ is connected, see Figure. 3. Thus,  the effective resistance  is given by equation (\ref{aa}). This is a variant of the Wu's two-point resistance of the cycle graph \cite{Wu}. Using the trigonometrical identity 
\begin{eqnarray}
\label{aaaaa}
\sin^2(l+1)n\pi/N=4\cos^2n\pi/N\sin^2l\pi/N+2\sin^2n\pi/N-2\sin^2nl\pi/N-\sin^2(l-1)n\pi/N,
\end{eqnarray}
 we may show that the effective resistance satisfy the following recursion
\begin{eqnarray}
\label{rec}
R(l+1)=l(1-l/N)+1/2(N-1/N)-2R(l)-R(l-1),
\end{eqnarray}
 where the first term in the above equation is nothing but the effective resistance of the $N$-cycle graph. From the above recursion, one may propose the following formula for the effective resistance
\begin{eqnarray}
\label{con}
R(l)=\left\{\begin{array}{cl}
\frac{l}{2}\Big(1-\frac{l}{2N}\Big)& \text{for }  l\ \text{even},\\
\frac{1}{4}\Big(N-\frac{l^2}{N}\Big)& \text{for }  \  l\ \text{odd}.
\end{array} \right.
\end{eqnarray}\ \
It is not difficult to check that our formula given by equation (\ref{con}) is correct, i.e, satisfy the above recursion formula given by equation (\ref{rec}). To do so, one has to take into a account that for $l$  even,  both $l+1$ and $l-1$ are odd, similarly for $l$  odd, both $l+1$ and $l-1$ are even. As a result, the effective  resistance of the connected graph  $C_{N}^{-2N}(1,2)$ is
\begin{eqnarray}
\label{conn}
R(l)=\frac{1}{N}\sum_{n=1}^{N-1}\frac{\sin^2nl\pi/N}{\sin^22n\pi/N}=\left\{\begin{array}{cl}
\frac{l}{2}\Big(1-\frac{l}{2N}\Big)& \text{for }  l\ \text{even},\\
\frac{1}{4}\Big(N-\frac{l^2}{N}\Big)& \text{for }  \  l\ \text{odd}.
\end{array} \right.
\end{eqnarray}
By using the above formula,  one can  easily check the symmetry enjoyed by the effective  resistance  $R(l)=R(N-l)$. This formula will play a crucial role in obtaining a closed formula for the effective resistance of the circular ladder.
\section{Discussion} \ \ To conclude, in this work, we were able to obtain  exact formulas for the effective resistance, the total effective resistance, the FPT  and  the MFPT of the simple random walk on the $N$-cycle with $4$-nearest neighbors, all these quantities are written in terms of the Fibonacci numbers. The techniques used in this paper are related to those given by the author in  \cite{Chair1}, in which we computed the effective resistance of the complete graph minus $N$ edges, where $N$ is odd.

\ \  The effective resistance was written in terms of  certain numbers that we called the Bejaia and the Pisa numbers. These numbers generalize the bisected Fibonacci and Lucas numbers. It was shown recently that  the effective resistance of a $2\times N$ resistor network, is written in terms of the Pisa numbers \footnote{Work in progress}.

\ \  We may check  that our formula for the effective resistance is correct using the Foster first identity \cite{Foster}, namely, $$ \sum_{i\sim j}R_{ij}=N-1,$$ where the sum is taken over all pairs of 
adjacent vertices. Due to the symmetry that we have for the graph $C_{N}(1,2)$,  the sum over all effective resistances that connect adjacent vertices may be written as $$  \sum_{i\sim j}R_{ij}= NR(1)+NR(2).$$
From our expression for the effective resistance given by equation (\ref{trig16}), and substituting $F_{1}=F_{2}=1$, then a simple computation gives indeed the Foster identity $  \sum_{i\sim j}R_{ij}=N-1$.

\ \  One might as well check that the hitting time  satisfies the set of equations \cite{Lovasz} given by $$ H_{i,j}= 1+\frac{1}{d(i)}\sum_{k\sim i}H_{k,j},$$
or equivalently,
$$\sum_{k\sim i}\big(H_{i,j}- H_{k,j}\big)=d(i).$$
For the $N$-cycle with $4$-nearest neighbors, set $ i=0$, $ j=1$ then the above  equation reduces to $3H_{0,1} -H_{0,2}-H_{0,3}=4$, that is, the degree for the graph $C_{N}(1,2)$, in obtaining this result we have used our equation for the hitting time, (\ref{first hit}) and the fact that $F_{1}=F_{2}=1, F_{3}=2$.

\ \  It is interesting to note that by identifying  the total effective resistance  $R( C_{N}(1,2))$ given in the equations (\ref{trig22}),  (\ref{trig23}) with the invariat quantity $$ N\sum_{n=1}^{N-1}\frac{1}{4\sin^2(n\pi/N)+4\sin^2(2n\pi/N)},$$  
then, we deduce the following identities, 
\begin{eqnarray}
\label{trig26}
\sum_{n=1}^{N-1}\frac{1}{\sin^2n\pi/N}=\frac{N^2-1}{3},
\end{eqnarray}
\begin{eqnarray}
\label{trig27}
\sum_{n=1}^{N-1}\frac{1}{(1+4\cos^2n\pi/N)}&=&\frac{1}{2}\big( F_{2N-1}+2N-1\big)\frac{ 1}{\sqrt{5}}\Bigg(\frac{1+(\frac{3-\sqrt{5}}{2})^{N}}{1-(\frac{3-\sqrt{5}}{2})^{N}}\Bigg)\nonumber\\&+&\frac{1}{2}\big(F_{N-1}^{2}-F_{N}^{2}-1\big)  \text{for $N$ even },
\end{eqnarray}
\begin{eqnarray}
\label{trig28}
\sum_{n=1}^{N-1}\frac{1}{(1+4\cos^2n\pi/N)}&=&-\frac{1}{2}\big( F_{2N-1}-2N+1\big)\frac{ 1}{\sqrt{5}}\Bigg(\frac{1-(\frac{3-\sqrt{5}}{2})^{N}}{1+(\frac{3-\sqrt{5}}{2})^{N}}\Bigg)\nonumber\\&+&\frac{1}{2}\big(F_{N}^{2}-F_{N-1}^{2}-1\big) \text{for $N$ odd }.
\end{eqnarray}
The first identity is well known, however,  we have not seen   last two identities in the literature.

\vspace{7mm}
{\bf Acknowledgment:}

I would like to thank the Abdus Salam International Centre for Theoretical Physics,Trieste, Italy for supports and
hospitality during Summer 2012.  
      
\newpage

\bibliographystyle{phaip}

 \begin{figure}[htp]
\centering
\includegraphics[scale=1.00]{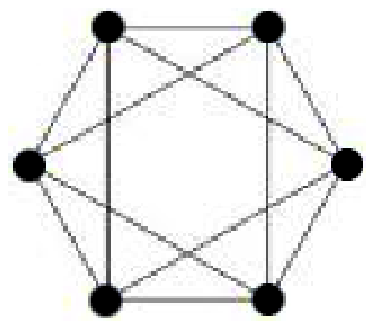}
\caption{Octahedral graph}
\label{Figure. 1}
\end{figure}
\begin{figure}[htp]
\centering
\includegraphics[scale=1.00]{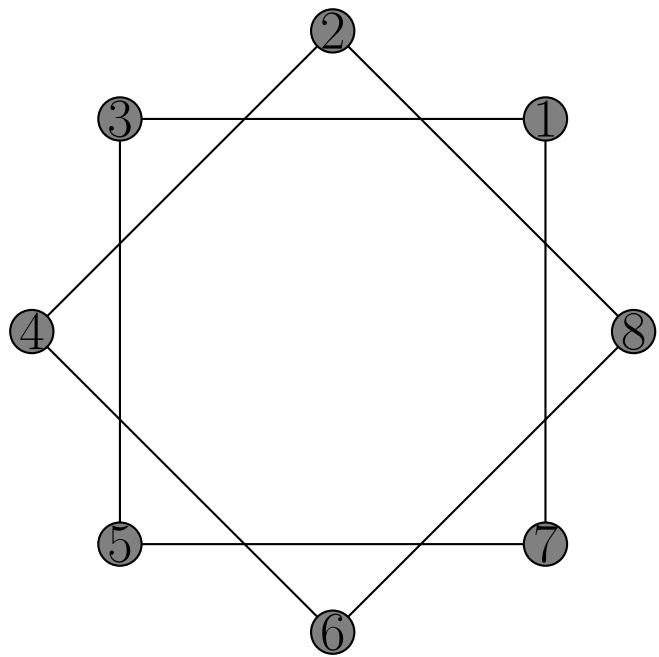}
\caption{Disconnected graph}
\label{Figure. 2}
\end{figure}
\begin{figure}[htp]
\centering
\includegraphics[scale=1.00]{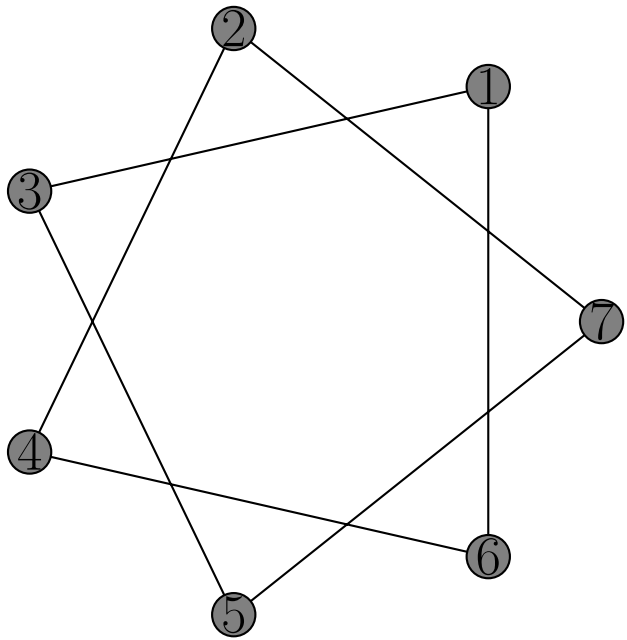}
\caption{Connected graph}
\label{Figure. 3}
\end{figure}
\end{document}